\documentclass[aps,prb,twocolumn,groupedaddress]{revtex4}

\usepackage{graphicx}
\usepackage{amssymb}

\newcommand{\rs}{\rm \scriptscriptstyle}

\begin{document}

\title{Three-body interactions with cold polar molecules}

\author{H.~P.\ B\"uchler, A. Micheli, and P. Zoller}
\affiliation{Institute for Theoretical Physics, University of Innsbruck,  6020 Innsbruck, Austria}
\affiliation{Institute for Quantum Optics and Quantum Information, 6020
Innsbruck, Austria}

\date{\today}

\begin{abstract}

We show that polar molecules driven by  microwave fields give naturally rise to
strong three-body interactions, while the two-particle interaction can be
independently controlled and even switched off. The derivation of these
effective interaction potentials is based on  a microscopic understanding of the
underlying molecular physics, and follows from a well controlled and systematic
expansion into many-body interaction terms.  For molecules trapped in an optical
lattice, we show that these interaction potentials give rise to Hubbard models
with strong nearest-neighbor two-body and three-body interaction. As an
illustration, we study the one-dimensional Bose-Hubbard model with dominant
three-body interaction and derive its phase diagram.

\end{abstract}


\maketitle

 Fundamental interactions between particles, such as the Coulomb law,
involves pairs of particles, and our understanding of the plethora of phenomena
in condensed matter physics rests on models involving effective two-body
interactions. On the other hand, exotic quantum phases, such as topological
phases or spin liquids, are often identified as ground states of Hamiltonians with three or more
body terms. While the study of these phases and properties of their
excitations  is presently one of the most exciting developments in
theoretical condensed matter physics, it is difficult to identify real physical
systems exhibiting such properties - a noticeable exception being the Fractional
Quantum Hall effect. Here we show that polar molecules in optical lattices
driven by microwave fields give naturally rise to Hubbard models with strong
nearest-neighbor three-body interactions, while the two-body terms can be tuned
(even switched off) with external fields.

The many-body Hamiltonians underlying condensed matter physics are derived
within an effective low energy theory, obtained by integrating out the high
energy excitations. In general, this gives rise to interaction terms
\begin{equation}
 V_{\rs eff}\left(\{{\bf r}_{i}\}\right)= \sum_{i< j} 
V\left({\bf r}_{i}\!-\! {\bf r}_{j}\right)+ 
\!\sum_{i<j< k}  W\left({\bf r}_{i},{\bf r}_{j},{\bf r}_{k}\right) +
\ldots  \label{effint}
\end{equation}
where $V({\bf r})$ describes the two-particle interaction depending only on the
separation between the particles. The second term  $W({\bf r}_{i},{\bf r}_{j},
{\bf r}_{k})$ is the three-body interaction, which depends on the distance and
orientation of three particles, and vanishes if one particle is far apart from
the other two. The ellipsis denotes possible higher many-body term terms.  While
for Helium atoms in the context of superfluidity the two-particle interaction
dominates and determines the ground state properties with the three-body
interactions providing small corrections,\cite{murphy71} model Hamiltonians with
strong three-body interactions have attracted a lot of interest in the search
for microscopic Hamiltonians exhibiting exotic ground state properties. Well
known examples are the fractional quantum Hall states described by the Pfaffian
wave functions   which  appear as ground states of a Hamiltonian with three-body
interaction. \cite{moore91,fradkin98,cooper04} These topological phases
admit anyonic excitations with non-abelian braiding statistic. Of special
interest are also  spin systems and bosonic Hamiltonians with complex many-body
interactions, such as ring exchange model, which are expected to give rise to
exotic phases.  \cite{moessner01,balents02,motrunich02,hermele04} Three-body
interactions are also an essential ingredient for systems with a low energy
degeneracy characterized by string nets,\cite{levin05,fidkowski06} which play an
important role in models for non-abelian topological phases.
The main challenge is then to identify experimental accessible systems, where
the two-particle interaction $V({\bf r})$ can be controlled, independent of the
three-body interaction $W({\bf r}_{i},{\bf r}_{j},{\bf r}_{k})$.

\begin{figure}[htb]
\begin{center}
\includegraphics[width=\columnwidth]{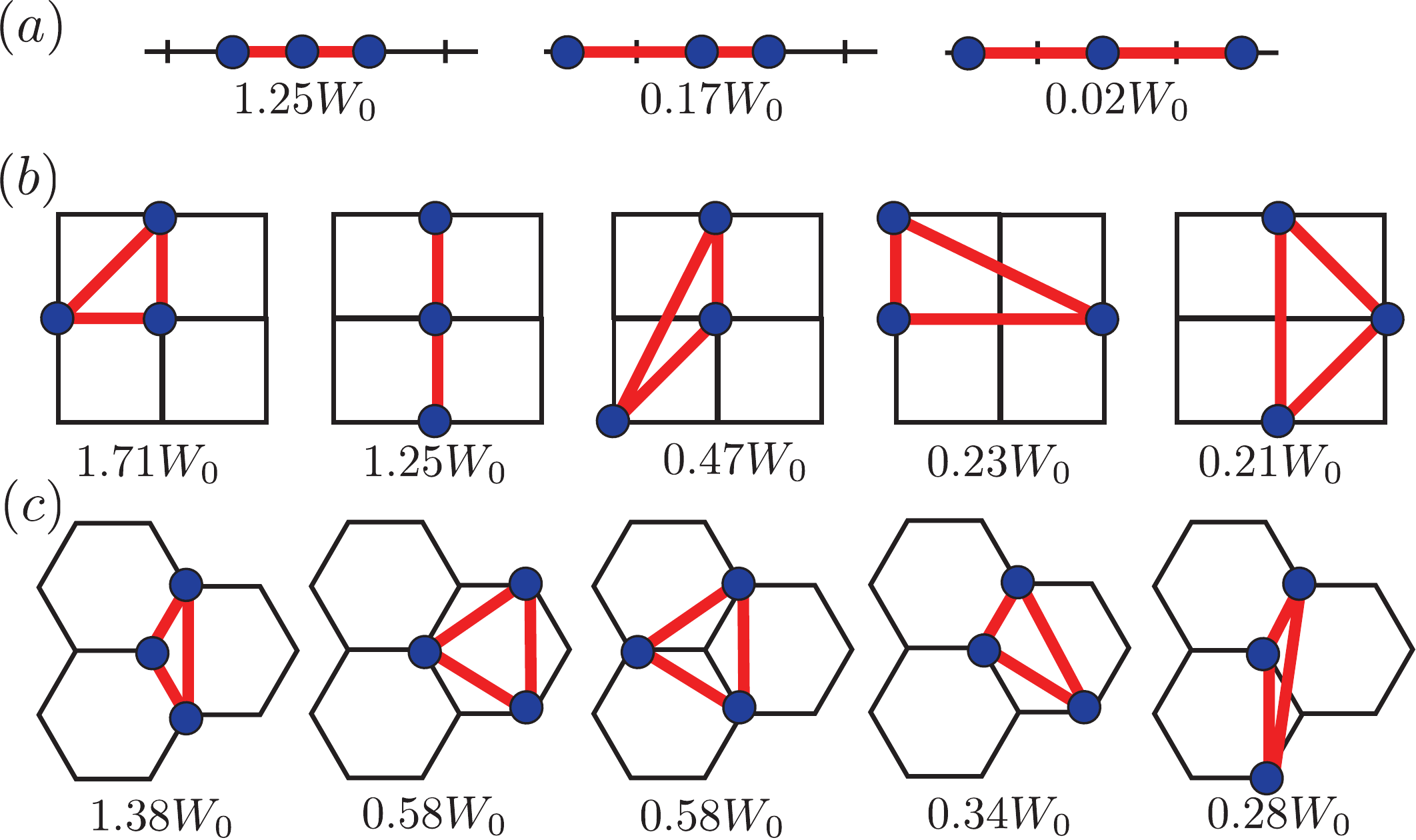}
\end{center}
\caption{ Strengths of the dominant three-body interactions $W_{i j k}$ appearing in the
Hubbard model for different lattice geometries: (a) one-dimensional setup (b)
two-dimensional square lattice (c) two-dimensional honey comb lattice.  The
characteristic energy scale $W_{0}= \gamma_{2} D
R_{0}^6/a^6$ is discussed following Eq.~(\ref{threebodystrength}).} \label{fig2}
\end{figure}

In the present work we analyze the effective interaction potential in a
many-body system of polar molecules. Recently, systems of polar molecules in the
rovibrational ground state have attracted a lot of interest recently due to
their rich internal structure and the presence of large permanent dipole
moments, which give rise to dipole-dipole interactions and offer the possibility
to  tune the interaction with static electric fields and microwave fields.
\cite{ronen06,kotochigova06,micheli06,baranov05,wang06,santos03} The techniques for
trapping and cooling of polar molecules with the goal to create quantum
degenerate  ground state molecules are currently developed in several
laboratories.
\cite{doyle04,sage05,rieger05,wang04,meerakker05,kraft06,sawyer07} We show below
that for an appropriate choice of static electric  and microwave fields  the
effective interaction reduces to the form as in Eq.~(\ref{effint}) with tunable
two-body and  the three-body interactions.

In particular, for polar molecules moving in an optical lattice we obtain the 
Hubbard model
\begin{equation}\label{Hubbard}
 H= - J \sum_{\langle i j \rangle} b^{\dag}_{i} b_{j} + \sum_{i \neq j}
\frac{U_{i j}}{2} n_{i} n_{j} + \sum_{i \neq j \neq k} \frac{W_{i j k}}{6} n_{i}
n_{j} n_{k}.
\end{equation}
Here $b_{i}$ ($b_{i}^\dagger$) are destruction (creation) operators for a
molecule on lattice site $i$, satisfying canonical commutation
(anti-commutation) relations for bosonic (fermionic) molecules,  and a hard-core
on-site repulsion is implied for bosons.  The first term in Eq.~(\ref{Hubbard})
is a hopping term (kinetic energy), while the last two terms describe two-body
and three-body interaction terms ($n_{i}=b_{i}^{\dag} b_{i}$). The different
strengths of the three-body interaction terms are illustrated in
Fig.~\ref{fig2}.  We emphasize that our derivation of the Hubbard model
Eq.~(\ref{Hubbard}), resulting in strong and tunable two and three-body
interactions, will be based directly on the effective many-particle potential
Eq.~(\ref{effint}). This is in contrast to the common approach to derive
effective many-body  terms from Hubbard models involving two-body interactions,
which are obtained in a $J\ll U$ perturbation theory, and are thus necessarily
small.\cite{tewari06} The main part of the present work is concerned with the
microscopic derivation of  the Hamiltonian in Eq.~(\ref{Hubbard}), and the
tunability of the parameters by external fields. As an illustration, we analyze
the simplest possible case of a one-dimensional Bose-Hubbard model with a
dominant three-body interaction, and derive the phase diagram using Bosonization
techniques.

\section{Effective interaction potential}

\begin{figure}[htb]
\begin{center}
\includegraphics[width=\columnwidth]{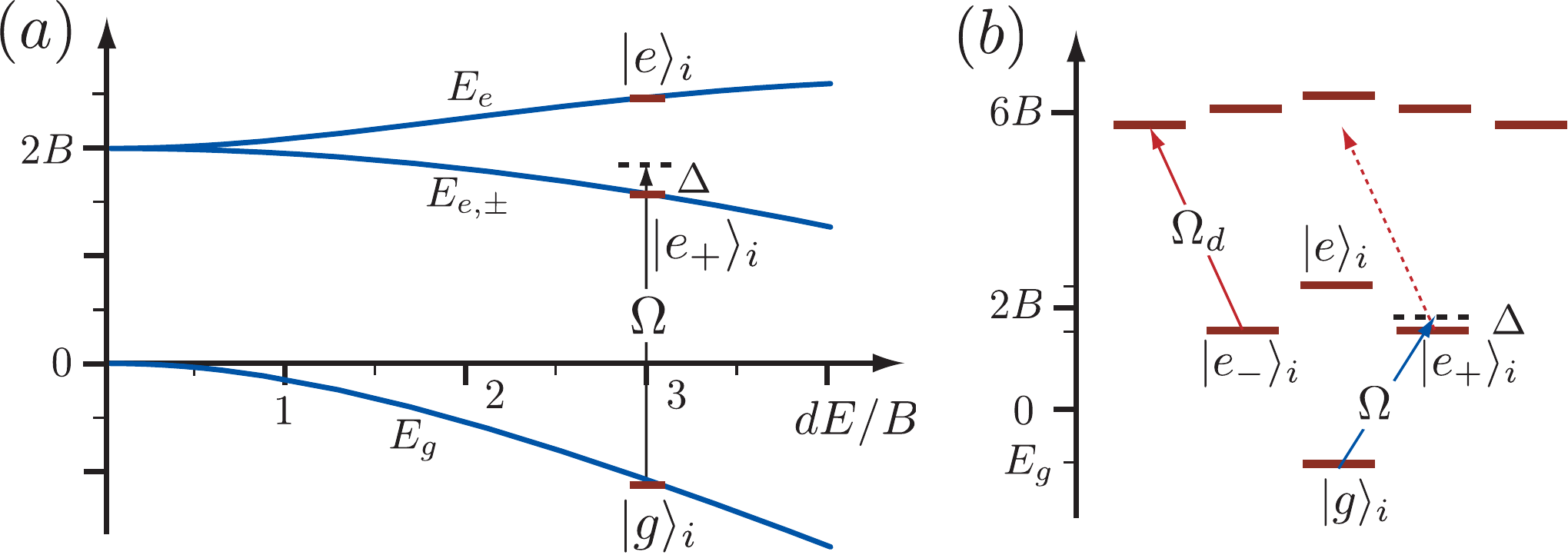}
\end{center}
\caption{(a) Internal excitation energies for a single polar molecule in a
static electric field ${\bf E}= E {\bf e}_{z}$. (b) Level structure for $E d/B =
3$: the circular polarized microwave field couples the ground state $|g\rangle$
with the excited state $|e_{+}\rangle$ with Rabi frequency $\Omega/\hbar$ and detuning
$\Delta$. Applying a second microwave field with opposite polarization (red
arrow) allows us to lift the degeneracy in the first excited manifold ($j=1$) by
resonantly couple the state $|e_{-}\rangle$ to the manifold with $j=2$. }
\label{fig0}
\end{figure}

The internal structure of polar molecules with a closed shell electronic
structure $ ^{1}\Sigma$  is given by the rotational degree of freedom, and its
low energy excitations are well described by the rigid rotor Hamiltonian
\begin{equation}\
 H_{\rs rot}^{(i)} = B {\bf J}_{i}^2
\end{equation}
with $B$ the rotational constant, and ${\bf J}_{i}$ the dimensionless angular
momentum operator. Here, we are mainly interested in the ground state
$|0,0\rangle_{i}$, and the first excited state manifold  $|1, m_{z}\rangle_{i}$
($|j, m_{z}\rangle_{i}$ denote the eigenstates of the rotor). The rotational
levels couple to external electric fields via $H^{(i)}_{\rs ext}= - {\bf d}_{i}
{\bf E}(t)$ with ${\bf d}_{i}$ the dipole operator. Under a static electric
field ${\bf E} = E {\bf e}_{z}$ along the $z$-axis the degeneracy of the excited
state manifold with $j=1$ is lifted, see Fig.~\ref{fig0}. In the following we
denote the new dressed states by $|0\rangle_{i} \rightarrow |g\rangle_{i}$ for
the ground state, $|1,0\rangle_{i} \rightarrow |e\rangle_{i}$ for the state with
$m_{z}=0$, and $|1,\pm 1\rangle_{i} \rightarrow |e_{ \pm}\rangle_{i}$ for the
remaining two degenerate states, and the corresponding energies $E_{g}$,
$E_{e}$, and $E_{e,\pm}$.

We focus on a setup with a circular polarized microwave field along the
$z$-axis, which couples the states $|g\rangle_{i}$ and $|e_{+}\rangle_{i}$ with
detuning $\Delta$ and Rabi frequency $\Omega/\hbar$.  We assume that the
degeneracy of the states $|e_{-}\rangle$ and $|e_{+}\rangle$ is lifted, e.g., by
an additional microwave field coupling the state $|e_{-}\rangle$ near-resonantly
to the next state manifold inducing an AC shift, see Fig.~\ref{fig0}; the
degenerate situation is presented in Appendix~A.  The internal
structure of a single polar molecule is then described as a spin-$1/2$ particle
via the identification of the state $|g\rangle_{i}$ ($|e_{+}\rangle_{i}$) as
eigenstates of the spin operator $S^{z}_{i}$ with positive (negative)
eigenvalue. In the rotating frame and applying the rotating wave approximation,
the Hamiltonian describing the internal dynamics of the polar molecule reduces
to
\begin{equation}
  H^{(i)}_{ 0} = \frac{1}{2} \left( \begin{array}{cc}
    \Delta & \Omega \\
     \Omega & - \Delta 
  \end{array} \right)
  = {\bf h} {\bf S}_{i}
\end{equation}
with the effective magnetic field ${\bf h}= (\Omega, 0, \Delta)$ and the
spin operator ${\bf S}_{i}= (S^{x}_{i},S^{y}_{i},S^{z}_{i})$.  
The eigenstates of this Hamiltonian are denoted as $|+\rangle_{i} =
\alpha |g\rangle_{i} + \beta |e_{+} \rangle_{i}$ and $|-\rangle_{i}= - \beta |g
\rangle_{i} + \alpha |e_{+} \rangle_{i} $ with energies $\pm \sqrt{\Delta^2 +
\Omega^2}/2$.
In the following, we consider a blue detuned microwave field with $\Delta >0$,
and molecule prepared into the state  $|+\rangle_{i}$. 
The preparation in this state is obtained by adiabatically turning on the
microwave field for molecules initially in their ground state $|g\rangle_{i}$.

The interaction between the polar molecules is determined by the dipole-dipole
interaction 
\begin{equation}
V_{\rs d-d}({\bf r}_{ij}) = \frac{{\bf d}_{i} {\bf d}_{j}}{|{\bf r}_{ij}|^3}-
\frac{
3 ({\bf r}_{ij}{\bf d}_{i})({\bf r}_{ij}{\bf d}_{j})}{|{\bf r}_{ij}|^5} 
\end{equation}
with ${\bf r}_{ij}={\bf r}_{i}-{\bf r}_{j}$ the separation between the
particles. In the interesting regime with  $|{\bf r}_{ij}|\gg  (D/B)^{1/3}$, the
dipole-dipole interaction restricted to the internal states $|e_{+}\rangle_{i}$
and $|g\rangle_{i}$ expressed in the rotating frame, and applying the rotating
wave approximation reduces to $H_{\rs d}=  H^{\rs ex}_{\rs d}+ H^{\rs dc}_{\rs
d}$. The first term takes the form  ($S^{\pm}_{i} = S_{i}^{x}\pm i S_{i}^{y}$)
\begin{equation}
 H^{\rs ex}_{\rs d}= -\frac{1}{2} \sum_{i \neq j} \frac{D}{2} \nu({\bf r}_{i j}) \: 
\left[S_{i}^{+} S_{j}^{-} + S_{j}^{+} S_{i}^{-}\right] 
\end{equation}
with the dipole coupling  $D= |\langle g| {\bf d}_{i} |e_{+} \rangle|^2$, while
the induced dipole moments $d_{g}= \partial_{E} E_{g}$  and $d_{e}= \partial_{E}
E_{e,+}$ gives rise to a second term
\begin{equation}
 H^{\rs dc}_{\rs d}= \frac{1}{2} \sum_{i \neq j} D \nu\left({\bf
r}_{ij}\right) 
\left[\eta_{g} P_{i} + \eta_{e} Q_{i} \right]\left[ \eta_{g}  P_{j}+ \eta_{e}
Q_{j}\right] \label{static}.
\end{equation}
Here, $P_{i}= 1/2+S_{i}^{z}$ and $Q_{i}=1/2-S_{i}^{z}$ are the projectors on the
ground and excited states, while  $\eta_{g} = d_{g}/\sqrt{D}$ and $\eta_{e}=
d_{e}/\sqrt{D}$  are the dipole couplings.  The
anisotropic behavior of the dipole-dipole interaction is accounted for by
$\nu({\bf r}) = (1-3 \cos^2\theta)/r^3$ with $\theta$ the angle between ${\bf
r}$ and the $z$-axis.

\begin{figure}[htb]
\begin{center}
\includegraphics[width=\columnwidth]{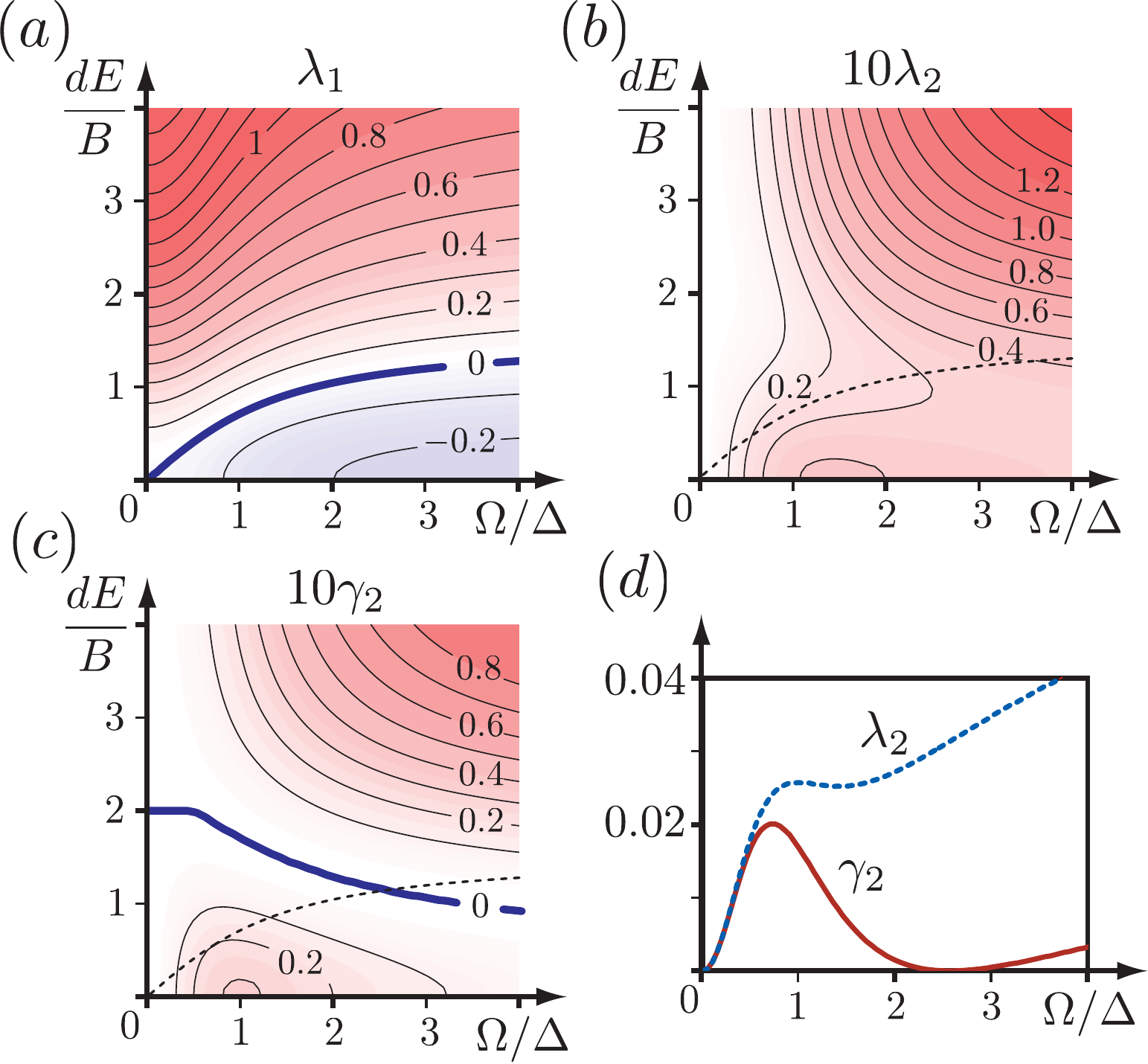}
\end{center}
\caption{(a)-(c) Strength of the interaction parameters $\lambda_{1}$,
$\lambda_{2}$, and $\gamma_{2}$ as a function of the external fields $E d/B$ and
$\Omega/\Delta$. The leading dipole-dipole interaction vanishes for
$\lambda_{1}=0$ (dashed line in (b) and (c)), and the second order contributions dominate the
interaction. (d) Strength of $\lambda_{2}$ (dashed line) and $\gamma_{2}$ (solid
line) along the line in parameter space with $\lambda_{1}=0$.} \label{fig1}
\end{figure}

Next, we are interested in the effective interaction between the polar molecules
with each molecule prepared in the state $|+\rangle_{i}$.  Within the
Born-Oppenheimer approximation, we determine the eigenenergies of the internal
Hamiltonian $\sum_{i}H_{0}^{(i)}+ H_{d}$ for fixed particle positions $\{{\bf
r}_{i}\}$, and obtain the energy shift of the state adiabatically connected to
the state $|G\rangle = \Pi_{i} |+\rangle_{i}$ of the non-interacting system. This energy shift is
driven by the dipole-dipole interaction $H_{\rs d}$ and strongly depends on the
positions of the particles $\{{\bf r}_{i}\}$ and therefore describes the
effective interaction $V_{\rs eff}(\{ {\bf r}_{i}\})$.  In the following, we
derive this energy shift using perturbation theory in the dipole-dipole
interaction $H_{\rs d}$, and derive the effective interaction potentials.  The
small parameter controlling the perturbative expansion is $D/(a^{3} |{\bf h}|)=
(R_{0}/a)^3$ with $a$ the characteristic length scale of the interparticle
separation, $|{\bf h}|=  \sqrt{\Delta^2+\Omega^2}$ the strength of the effective
magnetic field, and the length scale $R_{0}=(D/ \sqrt{\Delta^2+\Omega^2})^{1/3}$.
In first order perturbation theory, we obtain the energy correction
\begin{equation}
E^{(1)}(\{{\bf r}_{i}\})\! =\! \frac{1}{2} \left[ \left(\alpha^2 \eta_{g} \!+\!
\beta^{2} \eta_{e}\right)^2\! -\!  \alpha^2 \beta^2 \right] \sum_{i \neq j} D
\nu({\bf r}_{ij}), \label{firstorder}
\end{equation}
which describes a dipole-dipole interaction between the particles. In
addition, the energy shift in second order perturbation theory reduces to
\begin{eqnarray}
E^{(2)}\left(\{ {\bf r}_{i}\}\right)& = &
\sum_{k\neq i, k\neq j} \frac{\left| M \right|^2}{\sqrt{\Delta^2+ \Omega^2}} D^2
\nu\left({\bf r}_{i k}\right) \nu\left({\bf r}_{j k}\right)\nonumber\\
& & \hspace{-10pt}+ \sum_{i< j} 
\frac{\left| N \right|^2}{2 \sqrt{\Delta^2+\Omega^2}} 
 \left[D\nu\left({\bf r}_{i j}\right)\right]^2. \label{secondorder}
\end{eqnarray}
and gives rise to a correction to the two-particle interaction potential and an
additional three-body interaction.  The matrix elements $M$ and $N$ take the
form
\begin{eqnarray}
  M & = & \alpha \beta \left[ \left(\alpha^2 \eta_{g} + \beta^2
\eta_{e}\right)\left(\eta_{e} - \eta_{g}\right) + (\beta^2 - \alpha^2 )/2\right],
\nonumber \\ N & = & \alpha^2 \beta^2 \left[ \left(\eta_{e} - \eta_{g} \right)^2 + 1
\right].  \nonumber
\end{eqnarray}
Therefore, the effective interaction potential $V_{\rs eff}$ up to second order
in $(R_{0}/a)^{3}$ reduces to the form in Eq.~(\ref{effint}) with the
two-particle interaction potential
\begin{equation}
 V({\bf r}) = \lambda_{1} D \: \nu\left({\bf r}\right) + \lambda_{2}
D R_{0}^3 \;
\left[\nu\left({\bf r}\right)\right]^2, \label{twobody}
\end{equation}
and the three-body interaction
\begin{eqnarray}
 W\left({\bf r}_{1},{\bf r}_{2},{\bf r}_{3}\right) & =& \gamma_{2} R_{0}^3 D
\label{threebody} 
 \left[ \nu({\bf r}_{12})\nu({\bf r}_{13}) \right. \\  
& & \hspace{26pt}\left.+ \nu({\bf r}_{12})\nu({\bf r}_{23}) + \nu({\bf r}_{13})\nu({\bf r}_{23})
\right]. \nonumber
\end{eqnarray}
The dimensionless coupling parameters are $ \lambda_{1}  =
\left(\alpha^2 \eta_{g} + \beta^{2} \eta_{e}\right)^2 - \alpha^2 \beta^2$,
$\lambda_{2} = 2 |M|^2 + |N|^2/2$, and $\gamma_{2} = 2 |M|^2$.  These parameters
can be tuned via the strength of the electric field $Ed/B$ and the ratio between
the Rabi frequency and the detuning, $\Omega/\Delta$, see Fig.~\ref{fig1}.  Of
special interest are the values of the external fields, where $\lambda_{1}=0$,
i.e., the leading two-particle interaction vanishes.
Then, the interaction is dominated by the second order contribution with
$\lambda_{2}$ and $\gamma_{2}$, which includes the three-body interaction, see
Fig.~\ref{fig1}d, while small deviation away from the line $\lambda_{1}=0$
allows us to change the character of the two-particle interaction.
Note, that a $n$-body interaction term  ($n\geq 4$) appears in $(n-1)\mbox{-th}$
order perturbation theory in the small parameter $(R_{0}/a)^3$. Therefore, the
contribution of these terms is suppressed 
and can be safely ignored in our context.

The above analysis for $|G\rangle = \Pi_i |+\rangle$ ($\Delta>0$) provides a
positive energy shift in second order $E^{(2)}(\{{\bf r}_i\})$, as $|G\rangle$
corresponds to the highest energy state, and
provides a repulsive interaction with $\lambda_{2}\geq 0$ and $\gamma_{2}\geq0$. 
In turn, in an analogous analysis for the lowest
energy state $\Pi_i|-\rangle$, $E^{(2)}$ is negative, which yields a change
of sign of the coupling parameters $\lambda_2$ and $\gamma_2$.

The validity of the effective interaction $V_{\rs eff}(\{{\bf r}_{i}\})$ is
restricted to interparticle distances $|{\bf r}_{i}- {\bf r}_{j}|> R_{0}$.
Here, we prevent particles to approach each other on shorter distances $|{\bf
r}_{i}- {\bf r}_{j}|< R_{0}$ by focusing on setups where the combination of
interparticle potential for $|{\bf r}_{i}- {\bf r}_{j}|> R_{0}$ and the
transverse trapping potential produces a strong repulsive barrier.  For
sufficiently strong barrier height, thermal activation across the barrier and
quantum mechanical tunneling through the barrier are then suppressed, and the
particles are confined in parameter space to the region  $|{\bf r}_{i}- {\bf
r}_{j}|> R_{0}$. A setup providing such a barrier is obtained by confining the
particles into two-dimensions by a strong transverse trapping potential along
the $z$-axis with transverse trapping frequency $\omega_{\perp} = \hbar/m
a_{\perp}^2$ as provided for example by an optical lattice. The condition for an efficient
barrier in this two-dimensional setup has been recently worked out for polarized
molecules with leading dipole-dipole interaction. \cite{buechler07,micheli07}
However, the analysis can be generalized to the present situation with the
interaction potential $V_{\rs eff}(\{ {\bf r}_{i}\})$, if the two-particle
potential $V({\bf r})$ is sufficient repulsive, i.e., $\lambda_{1}\gtrsim -
\lambda_{2} (R_{0}/a)^3$.  For $\hbar \omega_{\perp} > D/R_{0}^3$, an estimate
of the rate for two-particles to penetrate the barrier up to a distance  $|{\bf
r}_{i}- {\bf r}_{j}|<R_{0}$ is  provided by $\Gamma \sim (\hbar/m a^2) \exp(- 2
S_{\rs E}/\hbar)$ with the semiclassical action $S_{\rs E}/\hbar \sim  \sqrt{D
m/R_{0} \hbar^2}$. This exponential suppression confines particle in parameter
space to the region $|{\bf r}_{i}- {\bf r}_{j}|> R_{0}$ for realistic parameters
with polar molecules, see below.

The low-energy many-body theory now follows by combining the kinetic energy of
the polar molecules with the  effective interaction $V_{\rs eff}$ within the
Born-Oppenheimer approximation  and the external trapping potentials $V_{\rs T}$
\begin{equation}
 H= \sum_{i }\frac{{\bf p}^{2}_{i}}{2 m} + V_{\rs eff}\left(\{ {\bf r}_{i}\}\right)
 + \sum_{i}  V_{\rs T}({\bf r}_{i}).
\label{lowenergyhamiltonian}
\end{equation}
Note, that the Hamiltonian is independent of the statistics of the particles and
therefore, valid for bosonic and fermionic polar molecules. 
In the strongly
interacting regime, where the interaction energy dominates over the kinetic
energy, it is expected that the ground state of the many-body system is determined
by crystalline phases. \cite{buechler07}

\section{Hubbard model}

Applying an optical lattice provides a periodic structure for the polar
molecules described by the Hamiltonian Eq.~(\ref{lowenergyhamiltonian}) and
allows us to derive Hubbard models with unconventional and strong
nearest-neighbor interactions.
We focus on the above setup, where the stability of the system is obtained by a
strong transverse trapping potential. We describe the lattice structure with
lattice spacing $a$ by a set of vectors $\{ {\bf R}_{i}\}$ accounting for the
minima of the optical lattice; depending on the optical lattice, we can generate
one-dimensional and two-dimensional systems. The mapping to the Hubbard model
follows the standard procedure:\cite{jaksch98} (i) Solving the single particle
problem in the presence of the optical lattice provides the Wannier functions
$w({\bf r})$ for the lowest Bloch band and determines the hopping energy $J$.
The Wannier functions describe localized wave functions with characteristic size
$a_{0}$.  (ii) We express the field operator $\psi$ in terms of the Wannier
functions in the lowest band $\psi^{\dag}({\bf r})= \sum_{i} w({\bf r}-{\bf
R}_{i}) b_{i}^{\dag}$ with $b_{i}^{\dag}$ ($b_{i}$) the creation (annihilation)
operator at the lattice site ${\bf R}_{i}$. In order to simplify the discussion,
we consider bosonic particles satisfying bosonic commutation relations.
Expressing the Hamiltonian  Eq.~(\ref{lowenergyhamiltonian}) in second
quantization and inserting the bosonic field operator $\psi$, maps the system to
the Bose-Hubbard model. However, the on-site interaction, which  derives for
cold atomic gases from the pseudopotential, requires a special discussion in the
present situation. In the experimentally interesting regime, we have the
following separation of the length scales:  $a_{0} \leq R_{0}< a$. With the
above discussion, that the parameter space of the system is confined to
interparticle distances larger than $ R_{0}$, this implies that if a particle is
at site ${\bf R}_{i}$, the hopping rate for a second particle to tunnel to this
site is strongly suppressed. As the initial system has no doubly occupied sites,
a convenient way to express this conditional hopping is to describe the bosons
as hard-core bosons.  Consequently,  no on-site interaction term is present, and
we obtain the Bose-Hubbard model in Eq.~(\ref{Hubbard}).
The interaction parameters $U_{i j}$ and $V_{i j k}$ derive from the effective
interaction $V_{\rs}\left( \{ {\bf r}_{i}\}\right)$, and in the limit of
well-localized Wannier functions ($a_{0}/a \ll 1$) reduce to $U_{i j} = V({\bf
R}_{i}- {\bf R}_{j})$ and $W_{i j k} = W({\bf R}_{i}, {\bf R}_{j}, {\bf
R}_{k})$.  The decay of these interactions with interparticle
separation takes the form
\begin{eqnarray}
U_{i j} & =& U_{0} \frac{ a^3}{|{\bf R}_{i}-{\bf R}_{j}|^3}+ 
U_{1}\frac{ a^6 }{|{\bf R}_{i}-{\bf R}_{j}|^6}, \\
 W_{i j k} & = & W_{0} \left[\frac{a^6}{|{\bf R}_{i}-{\bf R}_{j}|^3|{\bf
R}_{i}-{\bf R}_{k}|^3}+ perm\right].
\label{threebodystrength}
\end{eqnarray}
Here,  $U_{0} = \lambda_{1} D/a^3$, $U_{1}=
\lambda_{2} D R_{0}^3/a^6$, and $W_{0} = \gamma_{2} D R_{0}^3/a^6$ denote 
characteristic energy scales. The dominant contributions and strengths of 
the three-body terms in  different lattice geometries are shown in
Fig.~\ref{fig2}.  

In the following, we estimate these energy scales for  ${\rm Li Cs}$ with a
permanent dipole moment $d=6.3 {\rm Debye}$. Assuming an optical lattice with
lattice spacing $a\approx 500{\rm nm}$, the characteristic dimensionless
parameter determining the ratio between the interaction energy ($E_{\rs int}\!
=\! D/a^3$) and the characteristic kinetic energy within the lattice ($E_{\rs
kin}\! =\! \hbar^2/ m a^2$ proportional to the recoil energy) becomes $r_{d} = D m/\hbar^2 a
\approx 55$.  The leading dipole-dipole interaction can gives rise to very
strong nearest-neighbor interactions with $U_{0} \sim 55 E_{\rs kin}$. On the
other hand, tuning the parameters via the external fields to $\lambda_{1}=0$
the characteristic energy scale for the three-body interaction becomes $W_{0}
\approx (R_{0}/a)^3 E_{\rs kin}$.
Then, controlling the hopping energy $J$ via the strength of the optical lattice
allows us to enter the regime with dominant the three-body interactions.

\section{Three-body interactions in 1D}

As an illustration, we study the one-dimensional system with leading three-body
interaction. Tuning the external fields to a point with $U_{1} = -U_{0}$, 
the Hamiltonian takes the form
\begin{equation}
  H= -J \sum_{i} \left[b^{\dag}_{i} b^{}_{i+1} + 
  b^{\dag}_{i+1}b^{}_{i}\right] + W
\sum_{i} n_{i-1} n_{i} n_{i+1}, \label{1dthreebody}
\end{equation}
where we have kept only the nearest-neighbor terms. The additional interaction
terms play a minor role in the following analysis. Using a Jordan-Wigner
transformation allows us to map the hard-core bosons onto an interacting Fermi
system, which can be studied using Bosonization techniques;
\cite{gogolin98,sachdev99} the details of the calculation are presented in
Appendix B.  As a final result, we obtain that the low energy Hamiltonian
is described by the sine-Gordon model for the particle densities  $n=1/3$, $n=
1/2$, and $n = 2/3$,
\begin{equation}
  H = \frac{\hbar v}{2} \int dx \left[K \Pi^2+ \frac{1}{K}(\partial_{x}
\Phi)^2\right] + u \int dx \frac{1}{\pi^2 a^2} \cos\left( \beta \Phi \right).
\label{sineGordon2}
\end{equation}
The parameters $u$ and $\beta$ for the sine term  are given in Table
\ref{sineparameters}.  The renormalized Fermi velocity is $v=2 J a \sin(\pi n) /
\hbar K$, while the dimensionless Luttinger parameter takes the form
$K=1/\sqrt{1+\kappa}$ with
\begin{equation}
  \kappa = \frac{W}{2J} \frac{2 n}{\pi} \frac{3-2 \cos(2\pi n)- \cos(4 \pi n)
}{\sin(\pi n)}
\end{equation}
The Hamiltonian Eq.~(\ref{sineGordon2}) allows now for the derivation of the expected
phases diagram.
\begin{table}
\begin{tabular}{r|c|c|}
& \hspace{30pt}$u$ \hspace{30pt} &\hspace{30pt}$\beta$ \hspace{30pt} $\mbox{}$\\
\hline
$n=1/3$ & $W a/\pi$ & $\sqrt{36 \pi}$ \\ \hline
$n=1/2$ & $W a/2$ & $\sqrt{16 \pi}$ \\ \hline
$n=2/3$ & $ W a/\pi $ & $\sqrt{36 \pi} $ \\ \hline
\end{tabular}
\caption{Parameters of the sine-Gordon term \label{sineparameters}}
\end{table}

\begin{figure}[htb]
\begin{center}
\includegraphics[width=\columnwidth]{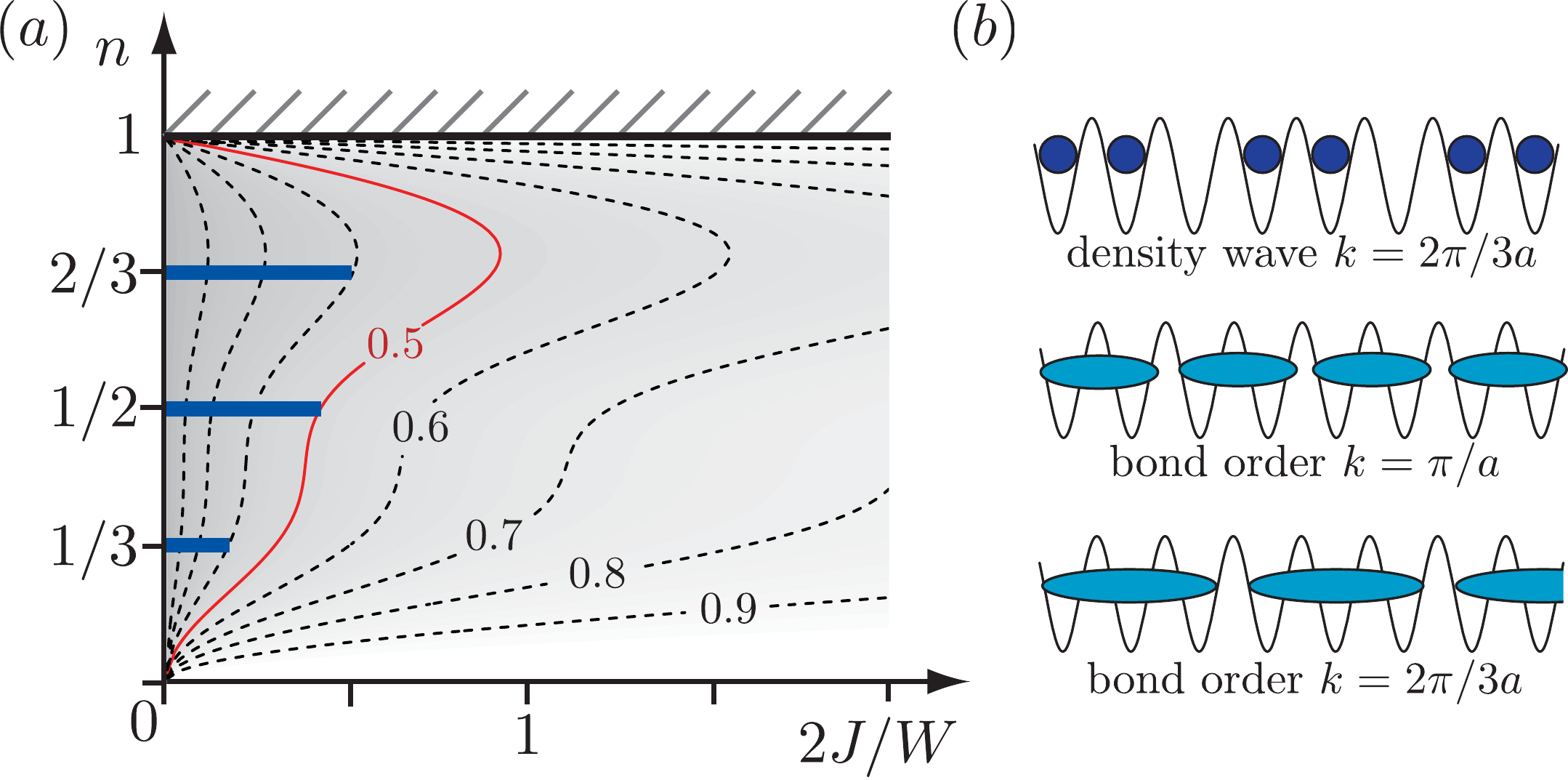}
\end{center}
\caption{(a) Sketch of the $W/J$-$n$ phase diagram: The dashed contours
correspond to constant values of $K$.  The first instability at $n=1/2$ appears
for $K=1/2$ (thin solid line), while the instabilities at $n=1/3$ and $n=2/3$
appear at lower values of $K$. Note that the interaction strength approaches
$W/2J \approx 1$ at the position of the instabilities, and we expect the exact
position of the instability to be strongly renormalized.  (b) Illustration of
the three different solid order characterizing  the gapped phases: density order
with wave vector $k=2 \pi /3a$, bond order with $k=\pi /a$, and bond order with
wave vector $k= 2 \pi /3 a$.
 } \label{fig3}
\end{figure}

For weak interactions $W/2J <1$, the three-body interaction shifts the hard-core
bosons away from the Tonks gas limit ($K=1$) into the correlated regime with $K<
1$. This regime is characterized by algebraic correlation functions, e.g., the
off-diagonal correlation between the bosons decays as $ \langle b_{i}
b^{\dag}_{j}\rangle \sim |i - j|^{- 1/2 K}$, and the density-density correlation
$\langle (n_{i}-n)(n_{j}-n) \rangle \sim \cos[2 \pi n (i-j)]/{|i -j|}^{2
K}$.\cite{haldane81}  At three values of the density $n = 2/3$, $n=1/2$,  and
$n=1/3$, the sine-Gordon term drives an instability towards a gapped phase, see
Fig.~\ref{fig3}. The critical value $K$ for this instability is given by $K=8
\pi/\beta^2$.
In the following, we use this criterion to identify the instabilities towards
three solid phases for $W/2J \sim 1$; the exact determination of the critical
interaction strength requires an analysis beyond the scope of this paper. These
solid phases are characterized by a broken translational symmetry, where the
following observables allow us to distinguish the different ground states
\begin{eqnarray}
 \langle n_{j} \rangle & \sim & \cos\left[ 2 \pi j  n + \sqrt{4 \pi} \Phi_{l} \right],
\label{densityorder}
\\
\langle b_{j}b_{j+1}^{\dag}+ b_{j}^{\dag}b_{j+1}\rangle & \sim & \cos\left[
(2\pi j + \pi) n+ \sqrt{4 \pi}\Phi_{l}\right] \label{bondorder} 
\end{eqnarray}
with $\Phi_{l}$ characterizing the different ground states.  The first
observable describes a density wave with wave vector $k = 2 \pi n/a$ present in
conventional solids, while the second observable accounts for a bond order with
wave vector $k = 2 \pi n/a$, see Fig.~\ref{fig3}b for an illustration  of the
different order appearing in the ground state.

At half filling $n=1/2$, the instability towards a gapped phase appears for $K =
1/2$. Within the gapped phase, the sine-Gordon term in Eq.~(\ref{sineGordon2})
determines the long-distance behavior, and the phase field $\Phi$ is
predominantly pinned within a minimum of the $\cos(\sqrt{16 \pi} \Phi)$ term.
The minima take the form  $\Phi_{l} = \pi ( 2 l+1)/\sqrt{16 \pi}$, and
characterize the different ground states in the gapped phase. These ground state
can be distinguished by the bond observable in Eq.~(\ref{bondorder}).
Therefore, we obtain a two-fold degenerate phase with a broken translational
symmetry: the bond correlation function exhibits a long range order at the wave
vector $k=\pi/a$, while the density $\langle n_{j} \rangle$ remains uniform in
this phase. The corresponding phase in spin systems is denoted as a spin-Peierls
phase.\cite{sachdev99}

In turn, for the densities $n=1/3$ and $n=2/3$ the instability appears at lower
values of $K$. The ground states of the gapped phase are then characterized by
the minima of $\cos(\sqrt{36 \pi} \Phi)$, which are given by $\Phi_{l}=\pi (2 l
+ 1)/\sqrt{36 \pi}$. The different ground states can be distinguished by the
density and  bond observable providing a three-fold degenerate phase with a
density wave and a bond order at the wave vector $k=2 \pi/3$. The appearance of
a density wave with  $k=2 \pi/3$ is a special property of the three-body
interaction.  It can be well understood for $n=2/3$ in the limit $W/2J \gg 1$:
then the ground state takes the form $\Pi_{i} b^{\dag}_{3 i}b^{\dag}_{3
i+1}|0\rangle$ and  describes a perfect density wave.

Within the above Bosonization approach, the additional interaction terms beyond
nearest-neighbor three-body interaction only provide a small renormalization of
the coupling parameters in Eq.~(\ref{sineGordon2}), and therefore play a minor
role in the qualitative discussion of the instabilities.
However, in the limit $W/2J\gg 1$ they become important and can give rise to
additional solid phases.

\section{Conclusions and Outlook}

A many-body system of cold polar molecules, whose rotational states are
dressed by an external static electric field and microwave field, 
can be described by an
effective Hamiltonian, where three-body interactions play the dominant role.
The derivation of this effective Hamiltonian involving interaction
potentials as a series of two-body, three-body etc. interactions is based
on a microscopic understanding of the underlying molecular physics, and is
systematic in the sense that the series expansion is well controlled. A
unique feature of the system is that, as a function of the external fields,
the two-body interactions can be tuned from repulsive to attractive, and
even switched off, while the three-body terms remain repulsive 
and strong. For molecules trapped in an optical lattice this leads to
Hubbard models with tunable nearest neighbor two-body interactions and
repulsive three-body terms. Models of this type have appeared in the recent
discussion of exotic quantum phases,  in particular in the context of
topological quantum phases and quantum computing, and we see molecular
quantum gases as a realistic experimental route, which provides the basic
building blocks and techniques towards the study of these phenomena.

\appendix

\section{ Degenerate states}

For a setup with $|e_{+}\rangle_{i}$ and $|e_{-}\rangle_{i}$ degenerate, 
it is necessary to keep the three states $|+\rangle_{i}$,  $|+\rangle_{i}$, 
and $|e_{-}\rangle_{i}$ for the perturbative calculation of the Born-Oppenheimer 
potentials. The leading contribution $E_{1}(\{ {\bf r}_{i}\})$ is not modified,
while the following term in the dipole-dipole interaction provides a
non-vanishing contribution in second order perturbation theory,
\begin{equation}
  \Delta H_{\rs d}^{\rs ex} = \frac{1}{2} \sum_{i\neq j}D  
\left[   \left[\mu({\bf r}_{i j})\right]^{*} T_{i}^{+} S_{j}^{-} + \mu({\bf r}_{i
j}) T_{i}^{-} S_{j}^{+}\right]
\end{equation}
with  the operators $T_{i}^{+} = |e_{-}\rangle \langle g|$  and $T_{i}^{+} = |g\rangle \langle
e_{-}|$ coupling the ground state with the excited state and the potential $
 \mu({\bf r}) =- 3 \left(x - i y\right)^2/2 r^5 $.
We therefore obtain a correction to the two-body interaction potential
\begin{equation}
\Delta V({\bf r}) = \lambda_{3} D R_{0}^3 \: \left| \mu({\bf r})\right|^2,
\end{equation}
and the three-body interaction
\begin{equation}
 W({\bf r}_{1}, {\bf r}_{2}, {\bf r}_{3}) = \gamma_{3}\frac{D R_{0}^3}{2}
\sum_{i \neq j \neq k}^{3}\left[\mu({\bf r}_{i k}) \right]^{*} \mu({\bf r}_{j
k}).
\end{equation}
The dimensionless parameters $\lambda_{3}$ and $\gamma_{3}$ take the form
\begin{eqnarray}
\lambda_{3} & = & 2 \beta^4 \alpha^2 \frac{\sqrt{\Omega^2+\Delta^2}}{\Delta+3
\sqrt{\Omega^2+\Delta^2}} +4 \alpha^4 \beta^2
\frac{\sqrt{\Omega^2+\Delta^2}}{\Delta+\sqrt{\Delta^2+\Omega^2}},\nonumber \\ 
\gamma_{3} &= & 4 \alpha^4 \beta^2
\frac{\sqrt{\Omega^2+\Delta^2}}{\Delta+\sqrt{\Delta^2+\Omega^2}},\nonumber
\end{eqnarray}
and depend on the external fields $E d/B$ and $\Omega/\Delta$.

\section{Bosonization} 
Using the equivalence between hard-core bosons and a spin-$1/2$ systems, allows
us to map the Hamiltonian in Eq.~(\ref{1dthreebody}) to a fermionic model with Fermi
operators $c_{i}$ ($c_{i}^{\dag}$) via a Jordan-Wigner transformation. 
Following the standard Bosonization procedure, we express the fermionic
operators via slowly varying left and right moving fields
\cite{gogolin98,sachdev99}
\begin{equation}   
c_{i} \sim  \sqrt{a} \left[e^{- i k_{\rs F} x_{i}} R(x)+ e^{i k_{\rs F}
x_{i}}
L(x)\right]
\end{equation}
with  $k_{\rs F}= \pi n /a$ the Fermi momentum and $n$ the averaged particle
density.  Here, $x_{i}$ describes the position of the lattice site $i$, while
$a$ denotes the lattice spacing. The fields $R(x)$ and $L(x)$ are slowly varying
and smooth on distances $a$ in the continuous variable $x \sim x_{i}$ with
\begin{eqnarray}
  R(x) & = & \frac{1}{ \sqrt{2 \pi a}} \exp\left(i \sqrt{4 \pi} \phi\right)\\
  L(x) & =&\frac{1}{ \sqrt{2 \pi a}} \exp\left(-i\sqrt{4 \pi} \bar{\phi}\right).
\end{eqnarray}
Then, the Hamiltonian for non-interacting fermions maps to the Luttinger liquid
Hamiltonian
\begin{equation}
   H_{0} = \frac{v_{\rs F}\hbar}{2} \int d{x} \left[\Pi^2 + (\partial_{x}
\Phi)^2 \right]
   \label{luttingerliquid}
\end{equation}
with the bosonic field $\Phi= \phi+\bar{\phi}$ and $\Pi$ the momentum conjugate
satisfying the canonical commutation relation $[\Phi(x),\Pi(x')]= i
\delta(x-x')$.  Here, $v_{\rs F}= 2 J \sin(\pi n)/\hbar$ denotes the Fermi
velocity of the non-interacting Fermi system.
In order to study the influence of the three-body interaction,
we split the density operator $n_{i}$ into its mean value and the fluctuations,
i.e., $n_{i} = n +\Delta n_{i}$.  Then, the interaction Hamiltonian reduces to
$H_{\rs int} = H_{\rs 1} + H_{2}+ H_{\rs 3}$ with
\begin{eqnarray}
   H_{ 2}& =& W n \sum_{i}\left[2 \Delta n_{i} \Delta n_{i+1}+ \Delta n_{\rs
  i-1}\Delta n_{\rs i+1}\right],\\
   H_{3} & = & W \sum_{i} \Delta n_{i-1} \Delta n_{i} \Delta n_{i+1},
\end{eqnarray}
while $H_{1}$ denotes a shift in the chemical potential and can be dropped.
The density operator expressed in the right and left moving fields takes the
form
\begin{displaymath}
 \Delta n_{i} = a \left[ \partial_{x} \Phi + e^{2 i k_{\rs F}x_{i}} M^{+}(x) +
e^{- 2
i k_{\rs F} x_{i}} M^{-}(x)\right]
\end{displaymath}
with
\begin{displaymath}
  M^{+}(x)= :R^{\dag}(x) L(x):, \hspace{10pt}  M^{-}(x)= :L^{\dag}(x) R(x):.
\end{displaymath}
Expressing the interaction Hamiltonian  $H_{\rs int}$ in terms of the bosonic
fields $\Phi$, special care has to be taken due to the  normal ordering  $:
\hspace{10pt}:$ of the operators. Furthermore, we note that some terms exhibit a
fast oscillations, which can be ignored in the low energy Hamiltonian.
Then, the interaction Hamiltonian $H_{2}$ takes the form
\begin{displaymath}
   H_{\rs int} = W n a\int dx \frac{1}{\pi}\left[ 3- 2\cos(2 \pi n)- \cos(4 \pi
n)\right]
\left(\partial_{x} \Phi\right)^2. 
\end{displaymath}
Furthermore, at half filling with $n=1/2$ an additional term appears due to
cancellation of different oscillating terms, which takes the sine-Gordon form
\begin{equation}
  H_{sG} =  u \int dx \frac{1}{\pi^2 a^2} \cos \left( \beta \Phi \right)
\label{sineGordon}
\end{equation}
with $u= W n a$ and $\beta= \sqrt{16 \pi }$.
The interaction $H_{3}$ only contributes less relevant terms, except at fillings
$n= 1/3$ and $n=2/3$. Then an additional terms appears, which takes the
sine-Gordon form in Eq.~(\ref{sineGordon}) with parameters $u=W a/\pi$ and $\beta =
\sqrt{36 \pi}$.

\acknowledgments

This work was supported by the Austrian Science Foundation (FWF),
the European Union projects OLAQUI (FP6-013501-OLAQUI), CONQUEST
(MRTN-CT-2003-505089), the SCALA network (IST- 15714), the
Institute for Quantum Information, and in part by the National Science Foundation under
Grant No. PHY05-51164.


\end{document}